%% file: main.tex
\documentclass[]{interact}

\usepackage{epstopdf}

\usepackage[natbibapa,nodoi]{apacite}
\renewcommand\bibliographytypesize{\fontsize{10}{12}\selectfont}

\usepackage{graphicx}
\usepackage{booktabs}
\usepackage{tabularx}
\usepackage{array}
\usepackage{ragged2e}
\usepackage{enumitem}
\usepackage{url}
\usepackage{xurl}
\usepackage[colorlinks=true,citecolor=blue,linkcolor=blue,urlcolor=blue]{hyperref}
\usepackage{orcidlink}

\newcolumntype{Y}{>{\RaggedRight\arraybackslash}X}
\setlist{noitemsep,topsep=2pt,partopsep=0pt}
\newif\ifanonymous
\anonymousfalse

\begin{document}

\articletype{RESEARCH ARTICLE}

\title{Comparing WeChat and Telegram MiniApp Security Architectures: Toward Cross-Platform Analysis}

\ifanonymous
  \author{
    \name{Anonymous Authors}
  }
  \hypersetup{pdfauthor={}}
\else
  \author{
    \name{
      Luca Ferrari\orcidlink{0009-0005-0209-6143}\textsuperscript{a},
      Billel Habbati\orcidlink{0009-0000-8315-7182}\textsuperscript{b} and
      Luca Verderame\orcidlink{0000-0001-7155-7429}\textsuperscript{b}
      \thanks{Corresponding Author: Luca Ferrari. Email: luca.ferrari@imtlucca.it}
    }
    \affil{
      \textsuperscript{a}IMT School for Advanced Studies Lucca, Lucca, Italy;
      \textsuperscript{b}University of Genoa, Genoa, Italy
    }
  }
  \hypersetup{
    pdfauthor={Luca Ferrari, Billel Habbati, Luca Verderame}
  }
\fi

\maketitle

\begin{abstract}
MiniApps are increasingly central to super-app ecosystems, yet architectural differences complicate the transfer of security analysis techniques across platforms. Existing WeChat analyzers rely on packaged applications and platform-specific APIs, limiting their direct applicability to Telegram’s web-based MiniApps. We present a qualitative comparison of WeChat Mini Programs and Telegram MiniApps, grounded in official documentation, prior research, and manual analysis of a specific Telegram Android source revision. Using a shared taxonomy of security controls, assets, and threat sources, we examine execution environments, communication paths, and trust boundaries. WeChat provides a dedicated MiniApp runtime; Telegram combines web execution with host controls. Both retain developer responsibilities for backend authentication and resource authorization. Two worked examples derive requirements for cross-platform analysis: distinguishing credential roles and validation boundaries, and associating host capabilities with concrete execution contexts and permission handlers. These findings inform future automated analysis without ranking the ecosystems by security or claiming validated vulnerability detection.
\end{abstract}


\section{Introduction}
\input{sections/01introduction}

\section{Related Work}
\input{sections/02relatedWorks}

\section{Methodology}
\label{sec:Methodology}
\input{sections/03Methodology}

\section{WeChat Mini Program Execution Ecosystem \& Communication}
\input{sections/03WeChatCommunication}

\section{Telegram MiniApp Execution Ecosystem \& Communication}
\input{sections/04TelegramCommunication}

\section{WeChat Mini Program Security Model}
\input{sections/05WeChatSecurityModel}

\section{Telegram MiniApp Security Model}
\input{sections/06TelegramSecurityModel}

\section{Discussion}
\label{sec:discussion}
\input{sections/07Discussion}

\section{Conclusion and Future Work}
\input{sections/08Conclusion}

\section*{Ethical considerations}

The study is a qualitative
comparison of two publicly documented MiniApp ecosystems and involved no human
subjects, personal data, or user information; therefore, no institutional review
board approval was required. The WeChat model was derived from official
developer documentation and peer-reviewed literature, while the Telegram model
was reconstructed through manual inspection of the publicly available Telegram
Android client source code. The analysis focused exclusively on architectural
components, communication paths, trust boundaries, and security mechanisms.
No third-party MiniApp, bot, backend service, or production system was accessed,
tested, or interfered with.

The paper discusses attack surfaces and security mechanisms at the architectural
level and does not disclose previously unknown vulnerabilities, exploit code, or
operational attack artifacts. The Telegram-specific findings are derived from
publicly available source code, while the WeChat characterization is based on
public documentation and prior literature. Should future empirical work identify
previously unknown exploitable weaknesses in specific applications or platform
components, we will follow responsible disclosure practices and notify the
affected parties before publication.

\section*{Declaration of AI usage}

During the preparation of this manuscript, the authors used ChatGPT
(OpenAI, GPT-5.5) and Grammarly (Grammarly Premium, web version accessed
in 2026) solely for language refinement, including grammar, spelling,
clarity, paraphrasing, and stylistic revision. The authors reviewed and
edited all AI-assisted suggestions, verified the originality and accuracy
of the manuscript, including the references, and take full responsibility
for its content. The authors also reviewed the applicable terms of use of
the employed tools and confirmed their suitability for this purpose.

\section*{Data availability statement}

This study uses official developer documentation, published literature, and the publicly available Telegram Android source code. Section~\ref{sec:Methodology} identifies the inspected revision and classes. Table~\ref{tab:evidence_mapping} provides source locations and permanent links to selected implementation paths, together with the architectural claims and analytical implications derived from them. Documentation and literature sources are cited in the manuscript.

The architectural diagrams, comparative taxonomy, and worked examples are presented in the paper. The examples are analytical illustrations, not runtime experiments. This study does not report a MiniApp corpus, instrumentation traces, or an evaluation of an automated detector.

\section*{Disclosure statement}
The authors report there are no competing interests to declare.

\section*{Funding}
This work was partially supported by the European Union -- NextGenerationEU under the Italian NRRP (MUR) program, project SERICS (PE00000014).

\section*{CRediT authorship contribution statement}
\textbf{Luca Ferrari}: Conceptualization, Methodology, Writing – original draft,  Writing – review \& editing. \textbf{Billel Habbati}:  Writing – review \& editing.
\textbf{Luca Verderame}: Conceptualization, Methodology, Writing – review \& editing,
Supervision.
\bibliographystyle{apacite}
\bibliography{software2}

\end{document}

%% file: sections/01introduction.tex
Mini Applications, hereafter MiniApps, have become an essential component of digital services in China, supporting applications ranging from mobile commerce to public administration. Before 2020, MiniApp ecosystems generally followed the App-in-App model \citep{MiniAppecosystem}, where the host application acts as a lightweight app store, enabling third-party developers to deploy apps within the host's environment. This model is largely platform-controlled and has been the focus of standardization efforts such as the W3C MiniApp Standardization White Paper \citep{MiniAppWhitePaper}.

Telegram introduced its own MiniApp model in 2022, targeting its base, which crossed 1 billion monthly active users in 2025 \citep{durov2025telegram1b}.
Telegram adopts a bot-oriented approach in which MiniApps are
associated with bot identities and delivered as web applications
built using standard web technologies. This differs from WeChat's
model, in which platform-specific application packages execute
within the host's MiniApp framework.
This architectural divergence motivates a systematic comparison of the differing trust boundaries and security guarantees across both platforms.

To avoid terminology ambiguity, we use \emph{MiniApp} as the generic cross-platform term throughout the paper. When discussing WeChat-specific artifacts or prior work that uses Tencent's naming, we use \emph{Mini Program}; when discussing Telegram, we use \emph{Telegram MiniApp}. This convention also separates the generic abstraction studied in this paper from platform-specific packaging and runtime names.

In this paper, we present a comparative study of WeChat Mini Programs and Telegram MiniApps, characterizing their architectural and security differences and establishing a basis for future automated vulnerability-detection methods. MiniApp ecosystems are rapidly expanding \citep{MiniAppecosystem}, and prior security analyses at scale \citep{meng2023wemint,wang2023taintmini,zhang2024minicat} rely on WeChat's package structure and client APIs, so they cannot be applied directly to Telegram's web execution model. By comparing WeChat's App-in-App paradigm with Telegram's bot-centric model, we identify the key abstractions that automated analysis frameworks must capture to support cross-platform reasoning.

Our contributions are as follows:
i) we provide a structured architectural comparison of WeChat's App-in-App model and Telegram's bot-centric web-based MiniApp framework;
ii) we analyze the security assumptions, trust boundaries, and attack surfaces of both ecosystems;
iii) we derive requirements for cross-platform security analysis and illustrate their application through two worked examples concerning credential roles and context-dependent permission mediation. The comparison is grounded in official documentation, prior literature, and manual inspection of the Telegram Android client, with selected claims linked to their supporting evidence in Table~\ref{tab:evidence_mapping}.

The rest of the paper is organized as follows. Section~\ref{sec:relatedworks} reviews related work. Section~\ref{sec:Methodology} outlines our comparative methodology. Section~\ref{sec:WeChatEcosystem} and Section~\ref{sec:TelegramEcosystem} detail the execution ecosystems and communication channels of WeChat and Telegram, respectively. Section~\ref{sec:WeChatSecurityModel} and Section~\ref{sec:TelegramSecurityModel} analyze their corresponding security models and attack surfaces. Section~\ref{sec:discussion} discusses analytical requirements through two worked examples, and Section~\ref{sec:conclusion} concludes the paper.

%% file: sections/02relatedWorks.tex
\label{sec:relatedworks}
Several papers have addressed MiniApp ecosystems, particularly in the WeChat domain. \citet{han2023systematic} identify nine major categories of issues: logic vulnerabilities, XSS, privilege escalation, SQL injection, arbitrary file uploads, weak password security, sensitive information leakage, CSRF, and insecure integration of third-party technologies. \citet{zhang2024minicat} propose MiniCAT, an automated analysis framework that crawls and unpacks WeChat Mini Programs and reconstructs attack paths across logic-layer JavaScript and render-layer WXML. \citet{yang2022cross} introduce Cross-MiniApp Request Forgery (CMRF), where a malicious MiniApp forges requests to another MiniApp by exploiting missing checks on the sender's appId (the platform-assigned identifier for a registered Mini Program, see Section~\ref{sec:WeChatSecurityModel}). \citet{meng2023wemint} present WEMINT, a static taint-analysis framework for detecting sensitive data leaks; in a large-scale evaluation over 20K suspicious MiniApps, they found that 7,585 apps, or 36.5\%, contained sensitive-data leaks. \citet{zhang2023don}, \citet{baskaran2023measuring}, \citet{wang2023taintmini}, and \citet{wang2023one} further show that WeChat security depends on subtle interactions among packaging, identity, APIs, and platform-version differences, consistent with related measurement work \citep{zhang2021measurement}.

Telegram MiniApps are less mature as a research target. The closest prior study to our work is \citet{Mohammadi}, who compares Telegram MiniApps and WeChat Mini Programs while focusing primarily on Telegram attack chains, TON/blockchain-specific risks, unproxied frontend--backend communication, insecure configuration practices, limited WebView isolation, and a Telegram-oriented mitigation framework based on QuickJS sandboxing, mandatory HTTPS, MLS encryption, and wallet verification. While their study catalogs empirical attacks and practical mitigations, it motivates Telegram security by citing known WeChat vulnerabilities from prior literature, without analyzing WeChat's architectural specifications or developer documentation.
More recently, \citet{ferrari2026telegapper} present TeleGapper, a black-box dynamic analysis framework for investigating the privacy practices of Telegram MiniApps. The framework captures runtime network traffic, identifies communications with third parties, and compares observed data flows with the applications' disclosed privacy information. Their evaluation of 278 working Telegram MiniApps highlights widespread discrepancies between actual data flows and privacy disclosures, providing empirical evidence of privacy and transparency issues in the Telegram MiniApp ecosystem.

Our contribution is a comparative analysis based on an architectural systematization of both ecosystems. We evaluate both platforms across shared dimensions, including execution environments, bridge communication channels, and credential custody, distinguishing the component that enforces a security control from the component responsible for using it correctly. This distinction separates platform properties, such as bridge-mediated access to host capabilities, from application properties, such as backend authorization and credential storage.

The comparative dimensions are shared, but the evidence sources are asymmetric: WeChat is characterized primarily through official documentation and prior studies, whereas Telegram additionally receives a manual source-code analysis of the official Android client. This inspection connects WebView configuration, concrete JavaScript bridge implementations, event dispatch, and navigation handling to the architectural boundaries summarized in Table~\ref{tab:evidence_mapping}. The worked examples in Section~\ref{sec:discussion} demonstrate how these architectural distinctions inform tooling requirements, focusing on structural modeling rather than empirical vulnerability discovery benchmarks.

%% file: sections/03Methodology.tex
We conducted a qualitative, threat-informed comparative analysis of the
WeChat and Telegram MiniApp ecosystems, combining documentation review
with a manual inspection of the Telegram Android client source code.

For \textbf{WeChat}, the architectural reconstruction and security model presented in Sections~\ref{sec:WeChatEcosystem} and~\ref{sec:WeChatSecurityModel} are grounded in Tencent's official developer documentation and peer-reviewed security literature on WeChat Mini Programs \citep[e.g.,][]{han2023systematic,meng2023wemint,wang2023taintmini,zhang2023don,zhang2024minicat}. We selected sources using a two-step protocol. First, we used official Tencent developer pages as authoritative sources for platform mechanisms such as package structure, cloud communication, authentication, permission checks, and publication controls. Second, we included peer-reviewed papers returned by searches for \emph{WeChat Mini Program security}, \emph{MiniApp privacy analysis}, \emph{taint analysis}, \emph{cross-MiniApp request forgery}, and \emph{MiniApp unpacking}, prioritizing studies that reported concrete vulnerabilities, tool implementations, or empirical datasets. We excluded blog posts and secondary reports unless they pointed to official documentation. Because the official WeChat Android client is proprietary and closed-source, direct source code inspection is not possible. Moreover, existing literature already characterizes its execution model, permission mechanisms, token lifecycle, and known vulnerability classes in depth. Consequently, our WeChat characterization synthesizes official documentation and peer-reviewed studies, establishing an architectural baseline without direct client reverse engineering.

For Telegram, whose MiniApp ecosystem is less extensively characterized in the academic literature, we manually inspected the source code of the official Telegram Android client\footnote{\url{https://github.com/DrKLO/Telegram}} to reconstruct the architecture of its MiniApp execution environment. The inspected repository was Telegram for Android, commit \texttt{62b56a07ca7e30e39f7fd00a6728d6bbd716ca1c}.
The analysis followed the code paths involved in the creation and configuration of MiniApp WebViews, JavaScript bridge registration, communication between MiniApps and the Telegram client, permission-sensitive SDK events, bot-related interactions, and navigation handling. The client-side inspection focuses on \texttt{BotWebViewContainer} and its nested classes \texttt{BotWebViewProxy}, \texttt{WebViewProxy}, and \texttt{MyWebView}; \texttt{ArticleViewer} and its nested \texttt{PageLayout}; and the separate \texttt{Browser} and \texttt{MessagesController} classes. Additional inspected classes include \texttt{BotWebViewSheet}, \texttt{ChatAttachAlertBotWebViewLayout}, \texttt{BotLocation}, \texttt{BotDownloads}, and \texttt{SetupEmojiStatusSheet}. In particular, we manually examined WebView configuration, bridge registration, event dispatch through \texttt{postEvent}, handlers for \texttt{web\_app\_request\_*} events, and multi-window navigation logic.

We manually traced these implementation paths and the relationships among the involved components to reconstruct the Telegram MiniApp execution model and identify its principal trust boundaries and security controls. The architectural representations reported in Figures~\ref{fig:TelegramEcosystem} and~\ref{fig:TelegramAttackSurface} combine these client-side observations with official Telegram documentation for platform protocols and backend interactions, and Android WebView documentation for the underlying execution and isolation model.

For both ecosystems, we subsequently applied a common comparative lens covering
the packaging and execution model, authentication and token lifecycle, API
restriction mechanisms, publication and review controls, backend communication
paths, sensitive-data handling, security controls, assets, and threat sources.

Table~\ref{tab:evidence_mapping} links selected architectural claims to identifiable source locations, the corresponding taxonomy elements, and their implications for analysis. Source-code entries refer to the inspected Telegram Android revision; documentation and literature entries identify the relevant sections. The final column details analytical implications derived from these sources. As such, the table serves as a traceability map for central architectural distinctions rather than an exhaustive feature inventory.
We annotate each ecosystem using three categories: \textbf{Security Controls (S)}, platform or developer mechanisms intended to prevent or limit misuse; \textbf{Threats and Threat
Sources (T)}, actors or conditions that can violate trust assumptions, isolation boundaries, or platform security policies; and \textbf{Assets (A)}, credentials, identifiers, data, and execution contexts whose compromise affects security or user privacy.

\begin{table*}[!tbp]
\centering
\caption{Evidence-to-claim mapping for selected architectural
distinctions and their implications for analysis.}
\label{tab:evidence_mapping} 

\small
\setlength{\tabcolsep}{4pt}
\renewcommand{\arraystretch}{1.15}

\begin{tabularx}{\textwidth}{
    @{}
    >{\raggedright\arraybackslash}p{0.23\textwidth}
    >{\raggedright\arraybackslash}p{0.25\textwidth}
    >{\raggedright\arraybackslash}p{0.23\textwidth}
    >{\raggedright\arraybackslash}X
    @{}
}
\toprule
\textbf{Claim} &
\textbf{Evidence and location} &
\textbf{Taxonomy and boundary} &
\textbf{Implication for analysis} \\
\midrule

\textbf{E1. WeChat application credentials and user-session
material have different roles.}
&
\citet{baskaran2023measuring},
Section~2 and Appendix~A.1.
&
WeChat A02--A05; S05 and S10.
Application-to-platform authentication and user-session
establishment.
&
Associate each credential use with its principal and endpoint;
model backend authorization separately.
\\
\addlinespace[5pt]

\textbf{E2. Telegram Bot API authentication and MiniApp
initialization-data validation protect different boundaries.}
&
Telegram Bot API:
\href{https://core.telegram.org/bots/api\#authorizing-your-bot}
{``Authorizing your bot''} and
\href{https://core.telegram.org/bots/api\#making-requests}
{``Making requests''};
Telegram Mini Apps \citep{TelegramBot}:
\href{https://core.telegram.org/bots/webapps\#validating-data-received-via-the-mini-app}
{``Validating data received via the Mini App''}.
&
Telegram A01--A02; S04.
Bot-to-platform requests and frontend-to-backend identity
information.
&
Distinguish the validation performed for each request from
application-specific resource authorization.
\\
\addlinespace[5pt]

\textbf{E3. The same JavaScript interface name can refer to
different bridge implementations.}
&
\href{https://github.com/DrKLO/Telegram/blob/62b56a07ca7e30e39f7fd00a6728d6bbd716ca1c/TMessagesProj/src/main/java/org/telegram/ui/web/BotWebViewContainer.java\#L488-L514}
{BVC, lines~488--514}
and
\href{https://github.com/DrKLO/Telegram/blob/62b56a07ca7e30e39f7fd00a6728d6bbd716ca1c/TMessagesProj/src/main/java/org/telegram/ui/web/BotWebViewContainer.java\#L3344-L3395}
{3344--3395}.
&
Telegram S06.
WebView-to-host bridge registration and event dispatch.
&
Resolve the container mode and concrete bridge object before
assigning host capabilities.
\\
\addlinespace[5pt]

\textbf{E4. Write-access requests use operation-specific checks.}
&
\href{https://github.com/DrKLO/Telegram/blob/62b56a07ca7e30e39f7fd00a6728d6bbd716ca1c/TMessagesProj/src/main/java/org/telegram/ui/web/BotWebViewContainer.java\#L1928-L1994}
{BVC, lines~1928--1994},
\nolinkurl{web_app_request_write_access} handler.
&
Telegram S06.
User consent and permission to send bot messages.
&
Represent the existing-authorization branch and the
user-confirmation branch; a bridge call alone does not imply
permission.
\\
\addlinespace[5pt]

\textbf{E5. New-window navigation follows a distinct creation
and routing path.}
&
\href{https://github.com/DrKLO/Telegram/blob/62b56a07ca7e30e39f7fd00a6728d6bbd716ca1c/TMessagesProj/src/main/java/org/telegram/ui/web/BotWebViewContainer.java\#L431-L440}
{BVC, lines~431--440},
\href{https://github.com/DrKLO/Telegram/blob/62b56a07ca7e30e39f7fd00a6728d6bbd716ca1c/TMessagesProj/src/main/java/org/telegram/ui/web/BotWebViewContainer.java\#L4445-L4522}
{4445--4522};
\href{https://github.com/DrKLO/Telegram/blob/62b56a07ca7e30e39f7fd00a6728d6bbd716ca1c/TMessagesProj/src/main/java/org/telegram/ui/ArticleViewer.java\#L14835-L14845}
{AV, lines~14835--14845}.
&
Telegram A03; S05--S06.
Navigation, destination context, and permission mediation.
&
Determine the destination WebView and registered handlers
before assessing capability inheritance.
\\
\addlinespace[5pt]

\textbf{E6. WeChat IP allowlisting is a configurable control.}
&
\citet{baskaran2023measuring},
Section~2, ``WeChat IP whitelisting''.
&
WeChat S04.
Supported server-side API access paths.
&
Record explicit deployment configuration, as platform support does not guarantee that the control is enabled.
\\

\bottomrule
\end{tabularx}

\vspace{4pt}

\begin{minipage}{\textwidth}
\footnotesize
\raggedright
\textit{Note.}
BVC denotes
\nolinkurl{TMessagesProj/src/main/java/org/telegram/ui/web/BotWebViewContainer.java};
AV denotes
\nolinkurl{TMessagesProj/src/main/java/org/telegram/ui/ArticleViewer.java}.
All code locations refer to commit
\nolinkurl{62b56a07ca7e30e39f7fd00a6728d6bbd716ca1c}.
Security-control and asset identifiers are local to the
corresponding ecosystem.
Code entries substantiate structural implementation properties, while literature-derived entries reflect the scope of their original studies.
\end{minipage}

\end{table*}

The evidence basis is asymmetric. The WeChat characterization
primarily relies on official documentation and prior peer-reviewed
literature, whereas the Telegram characterization additionally relies
on manual inspection of a specific revision of the Android client
source code. Accordingly, source-code-backed statements are scoped
to the inspected revision and should not be assumed to hold unchanged
across Telegram client versions or platforms. The study therefore
provides a qualitative architectural comparison rather than an empirical assessment of vulnerability prevalence.

%% file: sections/03WeChatCommunication.tex
\label{sec:WeChatEcosystem}
WeChat Mini Programs are lightweight software applications, accessed directly within the WeChat client, that are distributed as a single compressed package retrieved from the platform's registry. This package is archived in the \texttt{.wxapkg} format \citep{WeChatDirectoryStruct}.
A typical \texttt{wxapkg} file includes: i) A configuration file named \texttt{app.json}, which describes general features of the MiniApp (e.g., required permissions and entry points); ii) One or more folders containing resources such as images and audio files; iii) One or more folders defining the MiniApp's UI components, referred to as \emph{pages} by Tencent.
Each page is composed of four files:
\begin{enumerate}[leftmargin=*]
    \item A \texttt{.wxml} file, written in Tencent's proprietary WXML markup language, used to define UI structures such as buttons and input fields \citep{WeChatWXML}.
    \item A \texttt{.js} file, containing JavaScript logic executed in response to user interactions.
    \item A \texttt{.wxss} file, which specifies styling rules using WXSS, a variant of CSS customized for WeChat \citep{WeChatWXSS}.
    \item A \texttt{.json} configuration file, which defines UI behavior (e.g., window orientation and navigation settings).
\end{enumerate}

The above Mini Programs run in an architectural context involving multiple parties, described in the next section.
\subsection{WeChat Mini Program Execution Ecosystem}
WeChat Mini Programs run within a complex ecosystem involving multiple interacting entities. As shown in Figure~\ref{fig:ecosystem}, the main components of the WeChat Mini Program architecture include:
\begin{figure}[!ht]
    \centering
    \includegraphics[width=\columnwidth]{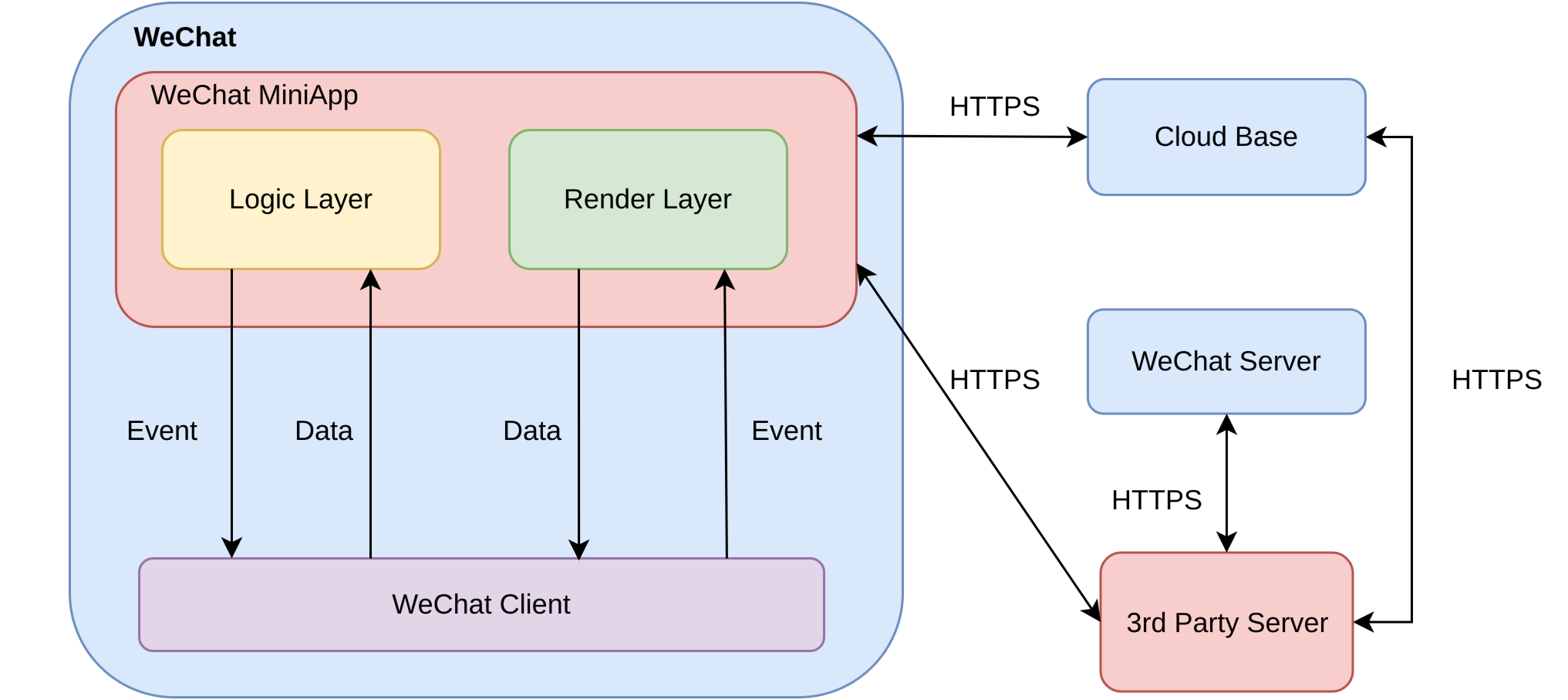}
    \caption{WeChat Mini Program Execution Ecosystem.}
    \label{fig:ecosystem}
\end{figure}

\noindent
\paragraph{WeChat Client}
The WeChat client is a super-app, installed on a user's device, hosting a collection of independent services and MiniApps within a single unified application. It provides the runtime environment for rendering, logic execution, and user interaction.

\noindent
\paragraph{WeChat Mini Program}
A WeChat Mini Program is a lightweight app whose pages are rendered in WebViews and whose scripts run in a separate JavaScript engine (JsCore), each managed by its own thread \citep{WeChatHostEnv}. It is conceptually divided into two distinct layers:

\begin{itemize}[leftmargin=*]
    \item \textbf{Render Layer:} This layer handles the user interface (UI) and runs the WXML templates and WXSS styles, with one WebView per page \citep{WeChatHostEnv}. WXML defines the structure of UI components and WXSS handles the styling. The render layer is responsible for visual presentation and captures user interaction, which it passes on to the logic layer.
    
    \item \textbf{Logic Layer:} This layer is responsible for managing user actions, data binding, and inter-page communication. It runs the MiniApp's JavaScript, including each page's \texttt{.js} file, and contains lifecycle methods, event-handling routines, and networking logic \citep{WeChatHostEnv}. It interfaces with the render layer through event propagation and data synchronization.
\end{itemize}

\noindent
\paragraph{WeChat Server}
The WeChat server manages MiniApp registration, authentication, and access control. It exposes server-side APIs (hosted at \texttt{api.weixin.qq.com}) that are called by the developer's backend rather than by the MiniApp, for example to obtain an \texttt{access\_token} or to exchange a login code for a session. These differ from the client-side APIs of the global \texttt{wx} object, which the MiniApp calls from the logic layer through the WeChat client.

\noindent
\paragraph{CloudBase}
As described by \citet{baskaran2023measuring} and \citet{wechatcloudbase}, CloudBase provides developers with native cloud support integrated into WeChat and reduces the need for backend and Operation \& Maintenance. Developers can implement core business logic through platform-managed APIs without provisioning dedicated servers, facilitating serverless deployment of Mini Programs and Mini Games while maintaining compatibility with external cloud infrastructure.

CloudBase has a range of features, e.g., cloud functions, databases, storage and cloud call. A cloud function allows developers to execute their server-side (JavaScript) code. 

These functions can be typically triggered by specific events, such as a user action or any change in data. Within the MiniApp, cloud functions can be triggered with the MiniApp's regular API (termed as JSAPI) \texttt{wx.cloud.callFunction}.

\noindent
\paragraph{Third-Party Server}
Third-Party Servers are used for advanced scenarios requiring integration with external systems or business-specific backend logic. WeChat supports IP allowlisting for server-side API access via the developer portal; because this protection requires explicit developer configuration \citep{baskaran2023measuring}, we classify it as optional hardening rather than mandatory platform enforcement.

\subsection{WeChat Mini Program Communication}
The communication within the Mini Program execution ecosystem involves several interactions: i) WeChat Mini Program \& WeChat Client, ii) WeChat Mini Program \& CloudBase, iii) WeChat Mini Program \& Third-Party Server, iv) Third-Party Server \& WeChat Server, and v) Third-Party Server \& CloudBase, as shown in Figure~\ref{fig:ecosystem}.

\noindent
\paragraph{WeChat Mini Program and WeChat Client}
The logic layer and render layer communicate with each other through an asynchronous and event-driven mechanism.

\noindent
\paragraph{WeChat Mini Program and CloudBase}
The communication between WeChat Mini Program and CloudBase happens through WeChat's internal API (e.g., \texttt{wx.cloud.callFunction} or \texttt{wx.cloud.database}) based on the HTTPS protocol.
This gives the MiniApp direct access to cloud functions without a Third-Party Server.

\noindent
\paragraph{WeChat Mini Program and Third-Party Server}
The WeChat SDK is responsible for the communication between the Mini Programs and the Third-Party Server through REST API communication via \texttt{wx.request()}.
This allows a MiniApp to send private data to a Third-Party Server for services unavailable in CloudBase.

\noindent
\paragraph{Third-Party Server and WeChat Server}
Communication between the two parties takes place through REST API calls. The goal is to obtain an \texttt{access\_token} \citep{WeChatAccessToken} by supplying the application's credentials, namely the \texttt{appID} and \texttt{AppSecret}, as parameters. Once retrieved, the \texttt{access\_token} authorizes the Third-Party Server to invoke the CloudBase API directly.

\noindent
\paragraph{Third-Party Server and CloudBase}
The Third-Party Server can directly perform REST API calls to the CloudBase service.

\begin{figure*}[!ht]
    \centering
    \includegraphics[width=0.6\textwidth]{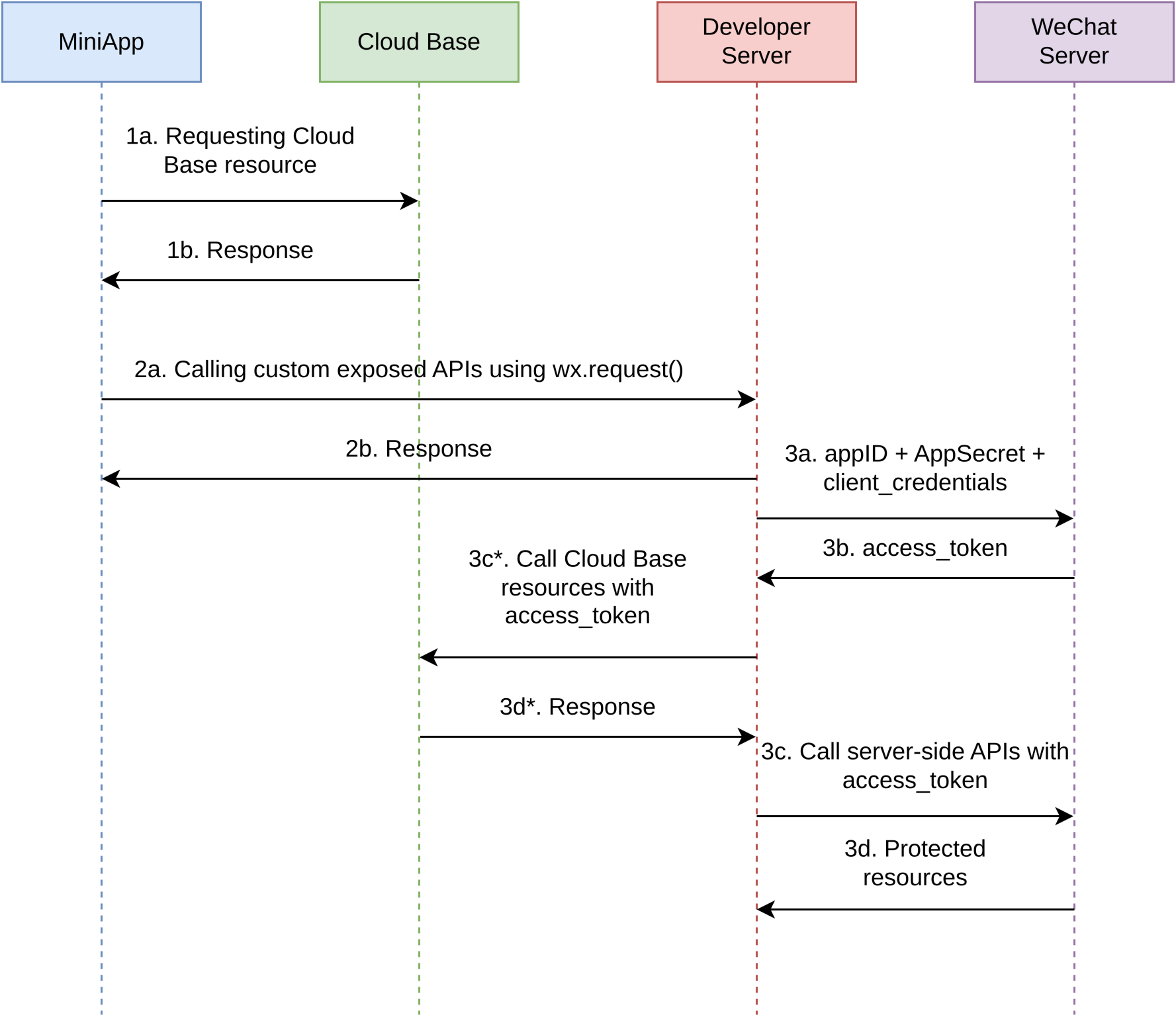}
    \caption{Communication between several entities of WeChat Mini Programs, synthesized from Tencent documentation and \citet{baskaran2023measuring}.}
    \label{fig:communicationWeChat}
\end{figure*}

Figure~\ref{fig:communicationWeChat} summarizes the principal communication paths. The illustration synthesizes the communication paths documented by \citet{wechatcloudbase} and \citet{WeChatAccessToken} and analyzed by \citet{baskaran2023measuring}.

%% file: sections/04TelegramCommunication.tex
\label{sec:TelegramEcosystem}

Telegram MiniApps differ from the packaged App-in-App model
used by WeChat and discussed in the W3C MiniApp Standardization
White Paper \citep{MiniAppWhitePaper}. In contrast to WeChat's packaged distribution model, Telegram MiniApps operate as standard web applications hosted on third-party servers (e.g., Netlify \citep{Netlify}). These apps are invoked via Telegram bots \citep{TelegramBot}, which act as entry points for their execution.

Due to the absence of a standardized packaging format or framework enforcement, the structure of a Telegram MiniApp is entirely dependent on the developer's choice of tools and build systems. Developers are free to implement their apps using raw HTML, CSS, and JavaScript, or popular web frameworks such as React \citep{React}.

\subsection{Telegram MiniApp Execution Ecosystem}

\begin{figure}[h]
    \centering
    \includegraphics[width=\columnwidth]
    {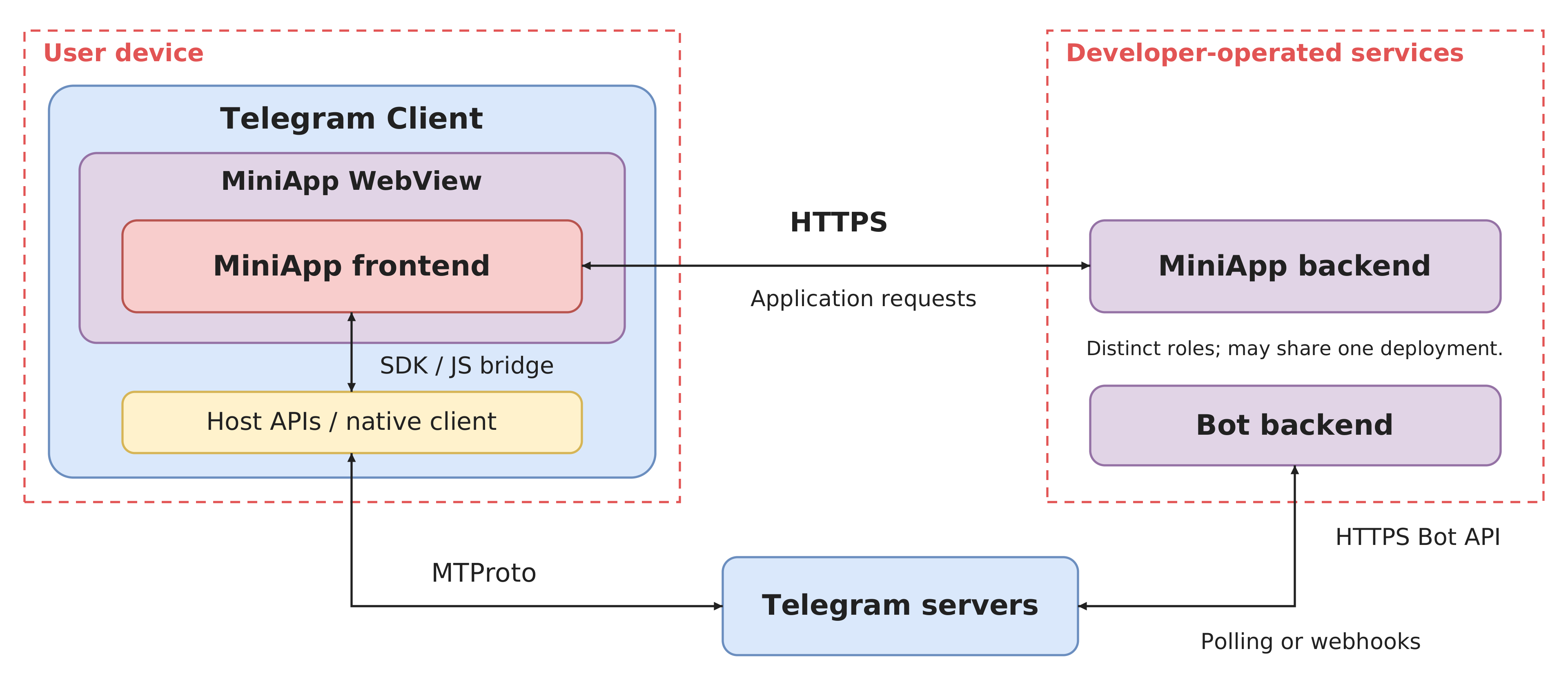}
    \caption{Telegram MiniApp execution ecosystem, showing
    client-side components, developer-operated backends,
    and Telegram servers.}
    \label{fig:TelegramEcosystem}
\end{figure}

Telegram MiniApps operate within a distributed execution ecosystem
comprising the Telegram client, developer-operated backends, and
Telegram servers. Figure~\ref{fig:TelegramEcosystem} illustrates
these components and their main communication paths.

\noindent
\paragraph{Telegram Client}
The Telegram Client is the host application installed on the
user's device. It loads and executes the MiniApp frontend inside
a WebView and mediates supported interactions with host
functionality through a local JavaScript bridge. The client
communicates with Telegram Servers using MTProto.

\noindent
\paragraph{Bot Backend}
The Bot Backend is a developer-operated service associated with
a Telegram bot identity. The bot's user interface is displayed
inside the Telegram Client, whereas its backend executes outside
the user's device. The backend invokes Telegram Bot API methods
over HTTPS and receives bot updates through either long polling
or a configured webhook endpoint.\footnote{\url{https://core.telegram.org/bots/api}}

\noindent
\paragraph{MiniApp Frontend and Backend}
The MiniApp Frontend executes inside the client's WebView and
uses the JavaScript SDK to request supported host interactions
through the local bridge \citep{TelegramBot}. It may also exchange
application data with a developer-operated MiniApp Backend over
HTTPS. The MiniApp Backend and Bot Backend are distinct logical
roles that may be implemented by the same service. Depending on
the application's functionality, the MiniApp Backend may also
invoke the Bot API. These calls are optional and require the
bot token to remain in trusted server-side components.

\noindent
\paragraph{Telegram Servers}
Telegram Servers provide the platform infrastructure accessed
by the native client through MTProto and by developer-operated
backends through the HTTPS Bot API. These are distinct
communication channels. Bot API requests are authenticated
using the bot token, whereas frontend interactions with the
client's SDK/bridge take place locally within the device.

%% file: sections/05WeChatSecurityModel.tex
\label{sec:WeChatSecurityModel}

\subsection{Attack Surface Overview}

\begin{figure*}[!h]
\centering
\includegraphics[width=0.76\textwidth]{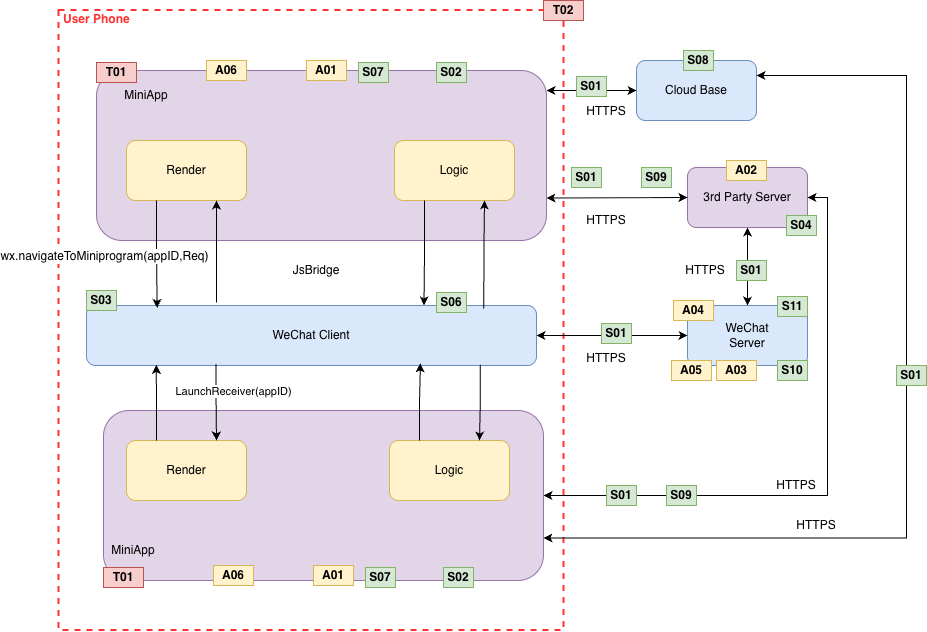}
\caption{WeChat Mini Program Attack Surface. Codes are defined in Table~\ref{tab:wechat-key}. The Third-Party Server represents the developer-operated MiniApp backend. \texttt{AppSecret} (A02) is associated with this server-side
component and must not be embedded in the MiniApp frontend.}
\label{fig:WeChatAttackSurface}
\end{figure*}

\begin{table}[!ht]
\centering
\footnotesize
\setlength{\tabcolsep}{4pt}
\renewcommand{\arraystretch}{1.1}
\caption{Key for Figure~\ref{fig:WeChatAttackSurface}: WeChat security controls, assets, and threat sources.}
\label{tab:wechat-key}
\begin{tabular}{@{}ll@{\hspace{1.2em}}ll@{}}
\toprule
\multicolumn{4}{@{}l}{\textit{Security controls}}\\
S01 & TLS & S07 & API Restrictions \\
S02 & Sandbox & S08 & Role-Based Access Control \\
S03 & Client Authentication & S09 & Data Encryption \\
S04 & IP Allowlisting & S10 & Credential Scoping \\
S05 & \texttt{access\_token} Authentication & S11 & Publication Review \\
S06 & Permission Mechanism & & \\
\addlinespace
\multicolumn{4}{@{}l}{\textit{Assets}}\\
A01 & \texttt{appID} & A04 & \texttt{session\_key} \\
A02 & \texttt{AppSecret} & A05 & Login Code \\
A03 & \texttt{access\_token} & A06 & User PII \\
\addlinespace
\multicolumn{4}{@{}l}{\textit{Threat sources}}\\
T01 & Malicious/Insecure MiniApp & T02 & Outdated Operating System \\
\bottomrule
\end{tabular}
\end{table}

The attack surface of WeChat Mini Programs spans the entities
introduced in Section~\ref{sec:WeChatEcosystem}: the WeChat
Client, MiniApp, WeChat Server, CloudBase, and Third-Party
Server, as illustrated in Figure~\ref{fig:WeChatAttackSurface}.
The principal execution boundary separates MiniApp code from
the host client's native capabilities. The logic and rendering
layers interact through platform-managed interfaces, while
access to host functionality is mediated by the MiniApp
framework and its bridge mechanisms.

Runtime isolation \textbf{(S02)} must be distinguished from
application-level authorization. Isolating MiniApp execution
does not establish that requests to developer-operated
backends are authorized or that application data is handled
securely. Likewise, rendering-process isolation should not
be interpreted as placing the entire WeChat client and all
MiniApp components within a single sandboxed WebView.

The effectiveness of runtime isolation depends on the client,
rendering engine, and underlying operating system. An outdated
operating system \textbf{(T02)} may retain vulnerabilities in
components on which these protections depend. Specific process
names and sandbox implementations are version-dependent and
are not treated here as universal properties of WeChat.

\subsection{Authentication and Access Control}

WeChat account authentication \textbf{(S03)} establishes access
to the host platform. It is distinct from the authentication
and authorization performed by an individual MiniApp backend.
An authenticated WeChat session does not, by itself, authorize
every operation on developer-operated services.

\begin{figure}[t]
\centering
\includegraphics[width=\columnwidth]{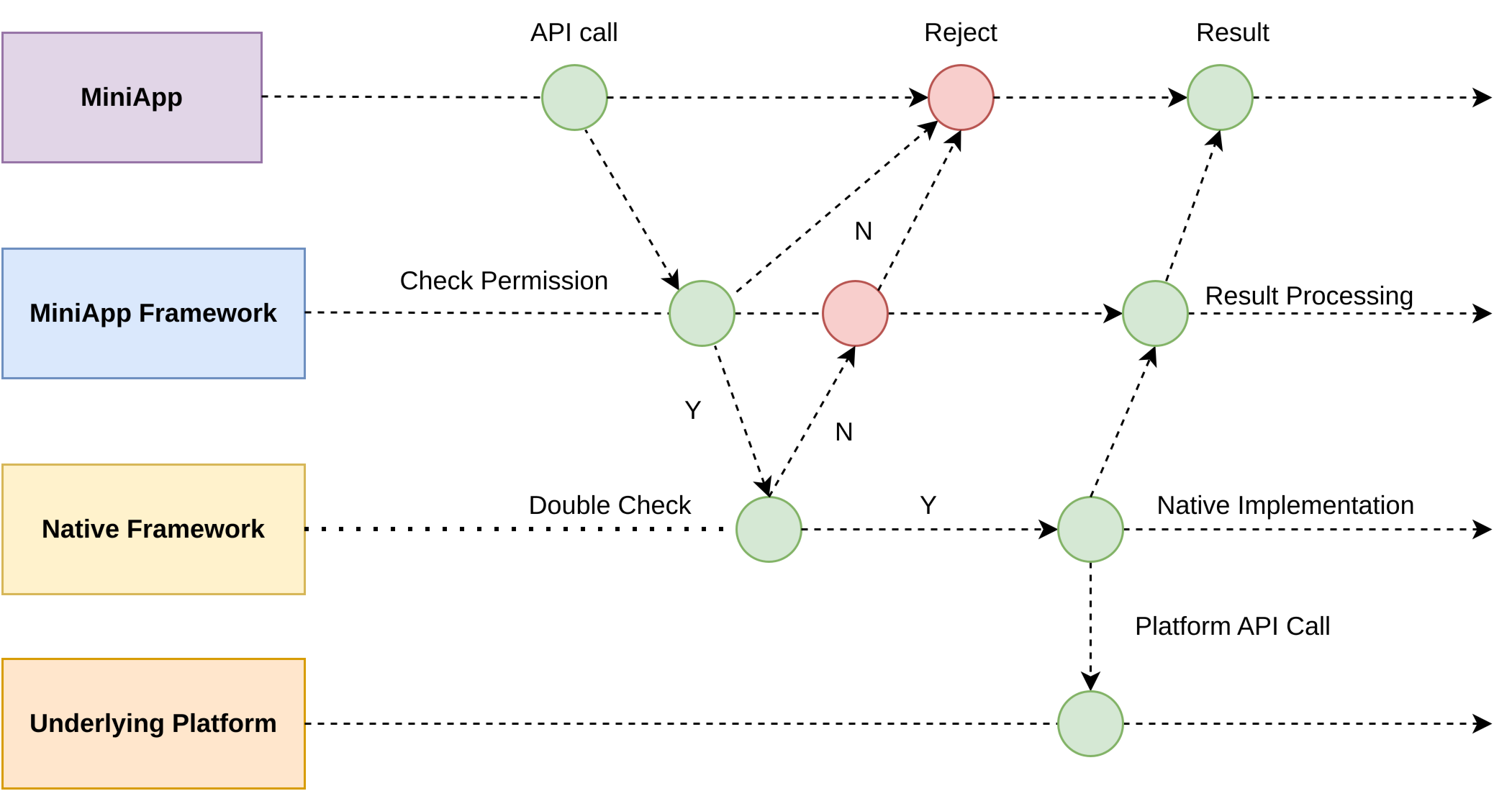}
\caption{Layered MiniApp permission model used in the WeChat comparison, adapted from \citet{MiniAppWhitePaper}.}
\label{fig:PermissionMechanism}
\end{figure}

Figure~\ref{fig:PermissionMechanism} presents the layered
permission model used to organize this comparison, adapted
from the MiniApp Standardization White Paper \citep{MiniAppWhitePaper}.
It distinguishes four layers: the \textbf{MiniApp}, which
requests an operation; the \textbf{MiniApp Framework}, which
mediates JavaScript API calls; the \textbf{Native Framework},
which implements access to native functionality; and the
\textbf{Underlying Platform}, which provides operating-system
capabilities and permissions.

At this level of abstraction, permission mediation
\textbf{(S06)} comprises the following steps:

\begin{enumerate}[leftmargin=*]
    \item The MiniApp requests a capability through a
    platform-provided API.
    \item The MiniApp framework evaluates the applicable
    permission requirements.
    \item The native implementation performs the relevant
    checks before invoking the underlying platform,
    subject to operating-system restrictions.
\end{enumerate}

This diagram represents conceptual layered mediation; concrete enforcement sequences vary depending on the requested capability, client version, and underlying operating system.

CloudBase introduces additional access-control boundaries
through account roles and resource permissions
\textbf{(S08)} \citep{wechatcloudbase}. Credential scoping
\textbf{(S10)} distinguishes application-level credentials
from user-bound session material. These controls should
not be interpreted as automatic enforcement of all
application-specific access policies.

Developers remain responsible for configuring cloud access
rules, validating inputs, enforcing resource ownership,
and handling returned data securely.

\subsection{MiniApp Architecture, Cloud Communication, and API Restrictions}

Building on the execution model introduced in
Section~\ref{sec:WeChatEcosystem}, communication with CloudBase
and platform APIs uses HTTPS \textbf{(S01)}. Transport
encryption protects data in transit, but does not prevent
a malicious or insecure MiniApp \textbf{(T01)} from sending
data to an endpoint that subsequently misuses it.

Developer-configured IP allowlisting \textbf{(S04)} can
provide an additional restriction on supported server-side
API access paths. We treat this mechanism as
platform-supported hardening whose applicability depends
on the API and configuration, rather than as a universal
restriction on all MiniApp network communication.

Server-side API authentication and user login involve
different credentials. In the former flow, the backend
uses the MiniApp's \texttt{appID} \textbf{(A01)} and
\texttt{AppSecret} \textbf{(A02)} to obtain an
\texttt{access\_token} \textbf{(A03, S05)}. In the latter,
the MiniApp obtains a short-lived login code
\textbf{(A05)} and forwards it to its backend, which exchanges
it together with the application credentials for user
identity and session material, including a
\texttt{session\_key} \textbf{(A04)}.

These flows cross different authentication boundaries.

An \texttt{access\_token} authorizes supported server-to-platform API operations, whereas a \texttt{session\_key} supports operations associated with
a particular user's session, including decryption in applicable encrypted-data flows. Neither replaces the MiniApp backend's own authorization checks.

Cross-MiniApp navigation is supported through platform APIs such as \texttt{wx.navigateToMiniProgram()}. Relevant parameters include the target \texttt{appId}, a \texttt{path} identifying a page within the target MiniApp, and \texttt{extraData} carrying additional data. This mechanism should not be equated with an arbitrary HTTP request to the target MiniApp.

The \texttt{appID} identifies an application and is not an
authentication secret. Cross-MiniApp interactions therefore
require validation of the sender information available
through the platform, the requested operation, and the
supplied data. Knowledge of a target identifier alone
does not establish authorization.

For APIs that return encrypted user data
\textbf{(A06)}, the backend uses the corresponding
\texttt{session\_key} \textbf{(A04)} to perform the
documented decryption procedure \textbf{(S09)}. This
description is API- and version-dependent: it does not
imply that every user-data retrieval flow uses
\texttt{session\_key} encryption.

The security consequences of credential exposure depend
on the credential and the remaining controls:

\begin{itemize}[leftmargin=*]
    \item Exposure of an \texttt{AppSecret} may permit
    unauthorized acquisition of an \texttt{access\_token},
    subject to applicable platform restrictions. Combined
    with a valid login code for the same MiniApp, it may
    also permit retrieval of the associated session
    material. Decrypting user data additionally requires
    access to the corresponding encrypted payload.

    \item Exposure of a login code may enable misuse of
    the login flow if an attacker can redeem or submit
    it before it becomes invalid and the backend fails
    to bind the resulting application session appropriately.
    Code exposure does not, by itself, demonstrate that
    one MiniApp can impersonate another.

    \item Exposure of an \texttt{access\_token} may permit
    unauthorized API operations within its scope and
    remaining validity period. The resulting impact
    depends on the accessible services and any additional
    restrictions.

    \item Exposure of a \texttt{session\_key} may compromise
    the confidentiality of corresponding encrypted user
    data. The key must therefore remain in trusted
    server-side components.
\end{itemize}

WeChat's MiniApp programming model constrains access to
APIs and rendering functionality \textbf{(S07)}. In
particular, application logic interacts with the rendering
layer through framework-provided mechanisms rather than
unrestricted direct DOM manipulation. Restrictions on
dynamic code evaluation must be interpreted in the context
of the execution environment and version being described.

Before publication, MiniApps are subject to WeChat's
platform review process \textbf{(S11)}. This distribution
control should not be interpreted as evidence of a
comprehensive manual security audit of every MiniApp or
as a guarantee that approved applications are free of
vulnerabilities.

\subsection{Assets Explanation}

\noindent
\paragraph{appID}
A public identifier assigned to a registered MiniApp.
It identifies the application in platform interactions
but does not independently authenticate a caller or
authorize access to application resources.

\paragraph{AppSecret}
An application secret used by trusted backend components to
authenticate supported requests to WeChat, including access-token
acquisition and login-code exchange. In the attack-surface diagram,
this credential (A02) is associated with the
\emph{Third-Party Server (MiniApp Backend)}, which represents the
developer-operated backend. The AppSecret must not be embedded
in MiniApp frontend code or distributed with the MiniApp package.
Its protection and rotation are developer responsibilities.

\paragraph{access\_token}
A time-limited credential used for supported
server-to-WeChat API calls. Backends must manage its
validity according to the relevant API response and
token-management requirements. It is an application-level
credential and must not be confused with a user login
code or an application-managed user session.

\paragraph{session\_key}
User-bound session material obtained by the backend
through the login-code exchange. In flows that return
encrypted user data, it supports server-side decryption.
The applicable data format, validity conditions, and
decryption procedure depend on the API and version.
The key should not be returned to the MiniApp frontend.

\paragraph{Login Code}
A short-lived credential obtained during the MiniApp
login flow and forwarded to the backend for exchange
with WeChat. It is distinct from an
\texttt{access\_token} and from the application session
subsequently established by the backend. Its limited
validity reduces the opportunity for misuse but does
not replace secure session binding and backend
authorization.

%% file: sections/06TelegramSecurityModel.tex
\label{sec:TelegramSecurityModel}

\subsection{Attack Surface Overview}
\begin{figure*}[!ht]
\centering
\includegraphics[width=\textwidth]{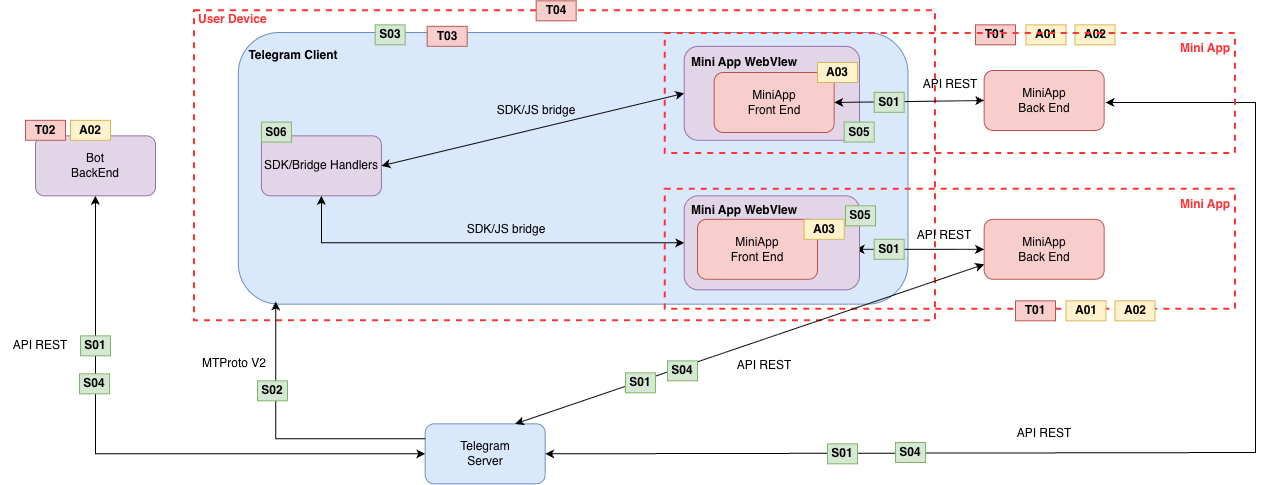}
\caption{Telegram MiniApp attack surface and principal
communication paths. MiniApp outlines denote logical
frontend--backend groupings. Bot API connections from MiniApp
Backends are optional. Codes are defined in Table~\ref{tab:telegram-key}.}

\label{fig:TelegramAttackSurface}
\end{figure*}

\begin{table}[!ht]
\centering
\footnotesize
\setlength{\tabcolsep}{4pt}
\renewcommand{\arraystretch}{1.1}
\caption{Key for Figure~\ref{fig:TelegramAttackSurface}: Telegram security controls, assets, and threat sources.}
\label{tab:telegram-key}
\begin{tabular}{@{}ll@{\hspace{1.2em}}ll@{}}
\toprule
\multicolumn{4}{@{}l}{\textit{Security controls}}\\
S01 & TLS & S04 & BOT\_TOKEN Authentication \\
S02 & Secure Protocol & S05 & Sandboxed Process \\
S03 & Client Authentication & S06 & Permission Mechanism \\
\addlinespace
\multicolumn{4}{@{}l}{\textit{Assets}}\\
A01 & initData & A03 & MiniApp Link \\
A02 & BOT\_TOKEN & & \\
\addlinespace
\multicolumn{4}{@{}l}{\textit{Threat sources}}\\
T01 & Malicious/Insecure MiniApp & T03 & Malicious/Insecure Telegram Client \\
T02 & Malicious/Insecure Bot & T04 & Outdated Operating System \\
\bottomrule
\end{tabular}
\end{table}

Telegram MiniApps operate through four core components (Figure~\ref{fig:TelegramAttackSurface}): i) \textbf{Telegram Client}, ii) \textbf{Telegram Server}, iii) \textbf{Bot}, and iv) \textbf{MiniApp (Web frontend + backend)}.
Each component contributes to the overall attack surface and introduces distinct security challenges.
\subsection{Telegram Client and WebView Configuration}
The Telegram client communicates with Telegram infrastructure using MTProto \textbf{(S02)}, while bot backends use the HTTPS Bot API \textbf{(S01)}. Separately, the client loads the MiniApp frontend into an Android WebView, which executes application JavaScript and exposes selected Telegram functionality through a registered bridge. These channels cross different trust boundaries and should not be represented as a single client-to-bot REST connection.

WebView configuration affects the resources and interfaces available to the frontend. Relevant properties include JavaScript execution, storage, file access, navigation handling, and bridge registration. Their interpretation must be scoped to the WebView instance and execution path under examination: settings applied to a general-purpose browser view cannot automatically be attributed to a bot MiniApp. Likewise, enabling a WebView feature does not establish that an application can use a protected device capability without additional permission checks.

Telegram account authentication \textbf{(S03)}, bridge-mediated permission decisions, and authorization at the MiniApp backend are separate controls. Being logged into Telegram does not remove the backend's responsibility to authenticate incoming application requests and authorize access to its resources.

\subsection{Bot Backend}

The Bot Backend authenticates its HTTPS Bot API requests using the \texttt{BOT\_TOKEN} (\textbf{A02}, \textbf{S04}), while TLS protects the transport channel (\textbf{S01}). Bot updates are received through long polling or webhooks. A MiniApp Backend may also issue Bot API requests, thereby performing part of the bot backend role. This explains the backend-to-Telegram connections in Figure~\ref{fig:TelegramAttackSurface}: they represent server-side bot operations whose presence depends on the deployment.

\subsection{MiniApp Execution and Risk Factors}

Telegram MiniApps consist of a web frontend and, where required,
developer-operated backend services. The frontend executes
inside a MiniApp WebView, while the backend runs outside the
user's device. The frontend--backend links shown in
Figure~\ref{fig:TelegramAttackSurface} use HTTPS
(\textbf{S01}). Transport protection does not prevent a
malicious or insecure MiniApp (\textbf{T01}) from mishandling
data at either endpoint.

A malicious or insecure bot (\textbf{T02}) introduces risks through its developer-operated backend and bot-authenticated interactions with Telegram. A malicious bot backend may misuse data it receives or perform unwanted actions within the bot's granted authority, whereas an insecure implementation may expose the \texttt{BOT\_TOKEN} (\textbf{A02}) or fail to enforce application-level authorization. Bot API authentication (\textbf{S04}) authenticates requests made using the bot's token; it does not establish the trustworthiness of the bot's implementation or authorize access to MiniApp backend resources.

The frontend executes within the isolation mechanisms provided
by the Android WebView environment (\textbf{S05}). Their
effectiveness depends on the client configuration, WebView
implementation, and underlying operating system. Standard
JavaScript and DOM operations, including dynamic code
evaluation where permitted by the configuration and application
policies, operate within this web execution context. They do
not independently grant access to protected native capabilities.
Frontend vulnerabilities such as cross-site scripting may
compromise application execution and expose accessible data,
including initialization data (\textbf{A01}).

The SDK/Bridge Handlers block represents client-side event dispatch
and permission mediation \textbf{(S06)}. In the inspected revision,
registration depends on the container's \texttt{bot} flag:
both branches use the JavaScript interface name
\texttt{TelegramWebviewProxy}, but bot contexts register
\texttt{BotWebViewProxy}, whereas non-bot contexts register
\texttt{WebViewProxy}. Their \texttt{postEvent} methods dispatch
to \texttt{onEventReceived} and \texttt{onWebEventReceived},
respectively (Table~\ref{tab:evidence_mapping}, E3).
Permission-sensitive bot events follow operation-specific handlers.
For example, \texttt{web\_app\_request\_write\_access} checks whether
the bot can already send messages and otherwise presents a
confirmation dialog before requesting permission (E4).
The interface name alone therefore does not identify the available
operations or their authorization requirements.

Multi-window navigation follows a separate path (E5).
The configuration enables \texttt{setSupportMultipleWindows(true)},
and \texttt{MyWebView.onCreateWindow} either obtains a
\texttt{WebView} through \texttt{ArticleViewer} or creates a
temporary \texttt{WebView} that intercepts the destination URL
and forwards it to \texttt{onOpenUri}.
\texttt{ArticleViewer.PageLayout} constructs its container with
\texttt{bot=false}; \texttt{onOpenUri} delegates URL handling
to \texttt{Browser.openUrl}.
These implementation relationships identify the contexts that
an analysis must distinguish. A new-window request alone does not
establish that the destination inherits the originating MiniApp's
capabilities or that a permission check has been bypassed.

The MiniApp Link (\textbf{A03}) is security-relevant because it determines the application entry point and the origin of the loaded content. Substitution of the intended link or compromise of its destination may direct users to attacker-controlled content. Opening the link does not by itself authenticate the user or authorize access to protected host functionality or backend resources; these operations remain subject to their respective permission and authorization checks.

A malicious or modified Telegram Client (\textbf{T03}) may
alter host-side restrictions and mediation mechanisms.
Such modifications change the assumptions under which the
controls operate and should be distinguished from vulnerabilities
in the official client. Similarly, an outdated operating system
(\textbf{T04}) may weaken the underlying protections on which
WebView execution depends.

\subsection{Assets Explanation}

\paragraph{BOT\_TOKEN} The \texttt{BOT\_TOKEN} authenticates bot requests to the Telegram Bot API. It belongs in trusted server-side components and must not be embedded in the MiniApp frontend. Exposure permits unauthorized actions within the bot's API authority.
The token also participates in the documented HMAC-based verification of initData; the security role of a credential therefore depends on the validation flow in which it is used.
It should be distinguished from user authentication data and from credentials protecting the MiniApp's own backend services.

\paragraph{MiniApp initialization data}
For launch modes that provide it, \texttt{Telegram.WebApp.initData} carries initialization data whose integrity can be verified by the backend. Depending on the launch mode and field availability, this data may include the user's Telegram identifier, first and last names, username, profile photo URL, language code, and Premium status. Theme settings are exposed separately through \texttt{Telegram.WebApp.themeParams} and are not part of the authenticated \texttt{initData} payload. Likewise, the user's IP address is not an \texttt{initData} field; servers may observe the source IP address of network requests independently of initialization data. The backend should verify the integrity of \texttt{initData} according to Telegram's documented procedure and check \texttt{auth\_date} against an appropriate freshness policy. The parsed \texttt{initDataUnsafe} object must not be trusted as authenticated input on its own \citep{TelegramBot}.

These checks establish a boundary between platform-provided identity information and application authorization. Valid initialization data does not automatically authorize access to every backend resource: resource ownership, permitted operations, and application session management remain backend responsibilities. The relevant distinction is therefore between Telegram account authentication, bot API authentication, and MiniApp backend authorization.

\noindent
\paragraph{MiniApp Link}
In this model, the MiniApp Link denotes the URL identifying an application entry point. It is treated as an asset whose destination integrity matters, rather than as an authentication credential.
\begin{table*}[!tbp]
\centering
\footnotesize
\caption{Comparison of MiniApp architectures and security responsibilities.}
\label{tab:miniapp-security-comparison}

\begingroup
\setlength{\tabcolsep}{5pt}
\renewcommand{\arraystretch}{1.0}

\begin{tabularx}{\textwidth}{
    @{}
    >{\raggedright\arraybackslash\bfseries}p{0.18\textwidth}
    >{\raggedright\arraybackslash}X
    >{\raggedright\arraybackslash}X
    @{}
}
\toprule Dimension & \textbf{WeChat Mini Programs} & \textbf{Telegram MiniApps} \\\midrule

Execution model
& Packaged MiniApps with separate logic and rendering layers.
& Remotely hosted web applications running in an Android WebView. \\
\addlinespace[2pt]

Isolation
& Platform-managed runtime boundaries.
& Android WebView isolation and host-client restrictions. \\
\addlinespace[2pt]

Backend communication
& CloudBase and developer-operated backend services.
&  HTTPS communication with developer-operated backends; server-side Bot API operations performed by the Bot Backend and, optionally, by the MiniApp Backend.\\
\addlinespace[2pt]

Permission checks
& Checks across the MiniApp framework, native implementation, and OS.
& SDK/bridge permission handlers and OS checks; separate external-navigation path. \\

\midrule

User authentication
& Backend exchanges a login code for user-bound session material.
& Backend validates \texttt{initData} before trusting user information. \\
\addlinespace[2pt]

Identifiers and credentials
& Public \texttt{appID}; secret \texttt{AppSecret};
  \texttt{access\_token}, \texttt{session\_key}, and login code.
& Server-side \texttt{BOT\_TOKEN}; verifiable \texttt{initData}
  and application-managed sessions. \\
\addlinespace[2pt]

Code and API access
& MiniApp-specific programming restrictions and platform APIs.
& Standard JavaScript and DOM functionality; selected host capabilities through the bridge. \\

\midrule

Publication review
& Platform review before MiniApp publication.
& Web deployment without an equivalent MiniApp package-review workflow. \\
\addlinespace[2pt]

Data protection
& Transport protection and API-specific encrypted-data flows.
& Transport protection; application-level data protection managed by developers. \\
\addlinespace[2pt]

Developer duties
& Protect credentials, validate inputs, enforce backend authorization, and configure platform services.
& Protect credentials, validate \texttt{initData}, and secure frontend and backend authorization. \\

\bottomrule
\end{tabularx}

\endgroup

\smallskip
\parbox{\textwidth}{\footnotesize
\textit{Scope:} The comparison reflects the documentation and
configurations examined in this study. Isolation depends on the
runtime and OS; publication review does not guarantee the absence
of vulnerabilities. In both ecosystems, authentication and
transport encryption do not replace application-specific
authorization or secure data storage.
}

\end{table*}

\begin{table*}[!tbp]
    \centering
    \caption{Comparison with \citet{Mohammadi} across shared security
    aspects and analytical scope.}
    \label{tab:comparison_mohammadi}

    \small
    \setlength{\tabcolsep}{5pt}
    \renewcommand{\arraystretch}{1.15}

    \begin{tabularx}{\textwidth}{
        @{}
        >{\raggedright\arraybackslash}p{0.21\textwidth}
        >{\raggedright\arraybackslash}X
        >{\raggedright\arraybackslash}X
        @{}
    }
        \toprule
        \textbf{Security aspect}
        & \textbf{\citet{Mohammadi}}
        & \textbf{This study} \\
        \midrule

        \textbf{Security responsibilities}
        &
        Contrasts WeChat's stronger platform-side enforcement
        with Telegram's greater reliance on developer practices.
        &
      Associates controls and credential uses with their validating components, distinguishing platform authentication from application-specific resource authorization.
        \\

        \addlinespace

        \textbf{Communication paths}
        &
        Discusses security and privacy risks associated with
        direct communication between Telegram MiniApps and
        external services.
        &
        Distinguishes frontend--backend communication,
        SDK/bridge interactions, and Bot API operations
        according to their endpoints and trust boundaries.
        \\

        \addlinespace

        \textbf{WebView execution and permissions}
        &
        Examines WebView-related weaknesses and exploitation
        scenarios in Telegram MiniApps.
        &
Links bridge registration, concrete event dispatchers, and navigation-created contexts to permission mediation in the inspected Telegram Android revision.
        \\

        \addlinespace

        \textbf{WeChat security mechanisms}
        &
        Discusses previously reported WeChat vulnerabilities
        as part of the ecosystem comparison.
        &
        Maps documented execution, authentication, permission,
        and publication mechanisms onto the same architectural
        categories used for Telegram.
        \\

        \addlinespace

        \textbf{Analysis objectives}
        &
        Investigates attack scenarios and proposes mitigation
        mechanisms, including measures addressing
        cryptocurrency-related threats.
        &
        Derives modeling requirements and illustrates them through two worked examples: credential-role analysis and context-dependent permission mediation. Detection effectiveness remains unevaluated.
        \\

        \bottomrule
    \end{tabularx}
\end{table*}

%% file: sections/07Discussion.tex
Table~\ref{tab:miniapp-security-comparison} compares the two ecosystems in terms of execution, communication, authentication, and security responsibilities. WeChat provides a MiniApp-specific runtime and platform services, whereas Telegram combines web execution with host-mediated
interactions and developer-operated infrastructure. These differences determine which components can enforce particular controls; they do not establish an unconditional security ranking.

Table~\ref{tab:miniapp-security-comparison} and Table~\ref{tab:comparison_mohammadi} synthesize the architectural comparison across identifiable components, credential roles, validation points, and execution contexts. Table~\ref{tab:evidence_mapping} connects these distinctions to supporting evidence. The following worked examples illustrate how this comparative model applies to concrete analysis tasks, showing how the taxonomy structures security reasoning without claiming empirical detection benchmarks.

\paragraph{Worked Example 1 (Credential Roles and Backend Authorization)}

Consider an analysis that encounters a credential in a MiniApp deployment. It must identify the credential's holder, the principal represented by its use, the receiving component,
and the operation being validated. In WeChat, application credentials and user-bound session
material participate in different flows (E1). In Telegram, bot-authenticated API calls and validation of MiniApp initialization data protect different boundaries (E2),
even though the \texttt{BOT\_TOKEN} participates in the documented HMAC-based validation procedure.
A unified analysis model must therefore contextualize each credential by its validating principal and endpoint instead of attributing a uniform security meaning to every token occurrence.

For an illustrative backend endpoint returning a user-owned record, establishing a user identity is only one step: the backend must also check whether that identity is
authorized to access the requested record. This yields a concrete analysis requirement: trace the authenticated identity to the resource-authorization decision,
separately from checks on platform credentials. If backend code or equivalent evidence is unavailable, the authorization property remains unassessed.
The presence of a platform token, transport encryption, or verified identity data cannot establish that property on its own.

\paragraph{Worked Example 2 (Navigation and Context-Dependent Permission Mediation)}

Consider a tool that flags \texttt{window.open()} as a permission bypass because the request does not pass through a MiniApp SDK handler. That rule omits the destination context. In the inspected Telegram revision,
\texttt{TelegramWebviewProxy} can denote either \texttt{BotWebViewProxy} or \texttt{WebViewProxy}, with different event dispatchers (E3).
New-window handling also follows a separate creation and routing path, including an \texttt{ArticleViewer} container configured in non-bot mode (E5).
A bridge name or navigation event is consequently insufficient to infer the destination's bot capabilities.

The corresponding analysis must associate each \texttt{WebView} with its creation path, context mode, concrete bridge implementation, and reachable handlers.
For a protected operation, it must then identify the applicable validation path; E4 provides a concrete example for write access.
A bypass claim additionally requires evidence that the protected operation is reachable without the required check. Static reconstruction identifies where to examine that
property, but does not demonstrate its violation or guarantee the security of every destination context.

\paragraph{Implications and Limits}

Together, these examples turn the taxonomy into two modeling requirements: distinguish the principal and validation boundary of each credential use, and associate host capabilities with concrete execution contexts and handlers.
Platform-specific adapters would be needed to recover these relationships from WeChat packages and APIs or Telegram web applications and client code.
The examples show how to formulate the analysis questions; whether this representation improves accuracy, coverage, or analyst agreement remains to be evaluated.

The evidence is asymmetric: WeChat relies on documentation and prior studies, while Telegram additionally uses source-code inspection of one Android revision.
The comparison does not measure vulnerability prevalence or establish behavior across client versions and platforms. Future evaluation should apply the model to independently
documented cases and test whether its classifications and derived checks remain useful across application designs and deployment configurations.

%% file: sections/08Conclusion.tex
\label{sec:conclusion}
This paper compares WeChat and Telegram MiniApps through a shared taxonomy of execution environments, communication paths, security controls, assets, and threat sources. The central distinction concerns the allocation of security responsibilities between the host platform and developer-operated components. Platform mediation, credential-based authentication, web execution, and backend authorization protect different boundaries and should be modeled separately.

The study contributes a qualitative architectural systematization supported by official documentation, prior work, and a manual inspection of the Telegram Android client source code. 
The source analysis supports the reconstruction of Telegram's execution contexts, bridge implementations, and permission-handling paths. Two worked examples translate these distinctions into requirements for credential-role analysis and context-dependent permission mediation. They illustrate the use of the taxonomy without establishing its effectiveness as a vulnerability-detection method.
Several directions follow from this systematization. First, the practical
utility of the proposed taxonomy for vulnerability discovery should be
evaluated empirically, for example by applying it to guide the analysis of a
corpus of real MiniApps. Second, the architectural relationships reconstructed
here from static sources could be validated dynamically, confirming which code
paths are exercised at runtime and under which client and device
configurations. Third, the comparison could be extended to other App-in-App ecosystems, such as Alipay \citep{AliPay} and Baidu \citep{Baidu}, evaluating whether these shared abstractions generalize beyond WeChat and Telegram. Finally, tooling that operationalizes
the taxonomy (mapping platform-specific mechanisms onto the shared controls,
assets, and threat categories) would help transfer these findings into
automated cross-platform analysis pipelines.

%% file: software2.bib
@misc{AliPay,
  author  = {{Alipay}},
  year    = {2026},
  title   = {Alipay},
  url     = {https://mobile.alipay.com},
  urldate = {May 25, 2026}
}

@misc{Baidu,
  author  = {{Baidu}},
  year    = {2026},
  title   = {Baidu},
  url     = {https://www.baidu.com/},
  urldate = {May 25, 2026}
}

@misc{wechatcloudbase,
  author  = {{Tencent}},
  year    = {2026},
  title   = {Weixin docs: CloudBase, getting started},
  url     = {https://developers.weixin.qq.com/minigame/en/dev/wxcloud/basis/getting-started.html},
  urldate = {May 26, 2026}
}

@misc{Netlify,
  author  = {{Netlify}},
  year    = {2026},
  title   = {Netlify: Push your ideas to the web},
  url     = {https://www.netlify.com/},
  urldate = {May 26, 2026}
}

@misc{React,
  author  = {{React}},
  year    = {2026},
  title   = {React: The library for web and native user interfaces},
  url     = {https://react.dev/},
  urldate = {May 26, 2026}
}

@misc{MiniAppWhitePaper,
  author  = {{W3C MiniApps Working Group}},
  year    = {2022},
  month   = oct,
  title   = {{MiniApp} standardization white paper, version 2},
  url     = {https://w3c.github.io/miniapp-white-paper/}
}

@article{zhang2021measurement,
  author  = {Zhang, Yue and Turkistani, Bayan and Yang, Allen Yuqing and Zuo, Chaoshun and Lin, Zhiqiang},
  year    = {2021},
  title   = {A measurement study of {WeChat} mini-apps},
  journal = {Proceedings of the {ACM} on Measurement and Analysis of Computing Systems},
  volume  = {5},
  number  = {2},
  pages   = {1--25}
}

@article{MiniAppecosystem,
  author  = {Schreieck, Maximilian and Ou, Ange and Krcmar, Helmut},
  year    = {2023},
  title   = {Mini-app ecosystems},
  journal = {Business \& Information Systems Engineering},
  volume  = {65},
  number  = {1},
  pages   = {85--93}
}

@inproceedings{baskaran2023measuring,
  author    = {Baskaran, Supraja and Zhao, Lianying and Mannan, Mohammad and Youssef, Amr},
  year      = {2023},
  title     = {Measuring the leakage and exploitability of authentication secrets in super-apps: The {WeChat} case},
  booktitle = {{Proceedings of the 26th International Symposium on Research in Attacks, Intrusions and Defenses}},
  pages     = {727--743}
}

@misc{WeChatDirectoryStruct,
  author  = {{Tencent}},
  year    = {2026},
  title   = {Weixin docs: Directory structure},
  url     = {https://developers.weixin.qq.com/miniprogram/en/dev/framework/structure.html},
  urldate = {July 21, 2026}
}

@misc{WeChatWXML,
  author  = {{Tencent}},
  year    = {2026},
  title   = {Weixin docs: {WXML}},
  url     = {https://developers.weixin.qq.com/miniprogram/en/dev/reference/wxml/},
  urldate = {July 21, 2026}
}

@misc{WeChatWXSS,
  author  = {{Tencent}},
  year    = {2026},
  title   = {Weixin docs: {WXSS}},
  url     = {https://developers.weixin.qq.com/miniprogram/en/dev/framework/view/wxss.html},
  urldate = {July 21, 2026}
}

@misc{WeChatHostEnv,
  author  = {{Tencent}},
  year    = {2026},
  title   = {Weixin docs: Mini Program host environment},
  url     = {https://developers.weixin.qq.com/miniprogram/en/dev/framework/quickstart/framework.html},
  urldate = {October 27, 2026}
}

@misc{WeChatAccessToken,
  author  = {{Tencent}},
  year    = {2026},
  title   = {Weixin docs: Getting access token},
  url     = {https://developers.weixin.qq.com/doc/offiaccount/en/Basic_Information/Get_access_token.html},
  urldate = {July 21, 2026}
}

@misc{TelegramBot,
  author  = {{Telegram}},
  year    = {2026},
  title   = {Telegram {Mini Apps}},
  url     = {https://core.telegram.org/bots/webapps},
  urldate = {May 26, 2026}
}

@misc{durov2025telegram1b,
  author  = {Mehta, Ivan},
  year    = {2025},
  month   = mar,
  title   = {Telegram founder {Pavel Durov} says app now has {1B} users, calls {WhatsApp} a ``cheap, watered down imitation''},
  url     = {https://techcrunch.com/2025/03/19/telegram-founder-pavel-durov-says-app-now-has-1b-users-calls-whatsapp-a-cheap-watered-down-imitation}
}

@misc{webview,
  author  = {{Google}},
  year    = {2026},
  title   = {{WebView}: {API} reference},
  url     = {https://developer.android.com/reference/android/webkit/WebView},
  urldate = {July 21, 2026}
}

@inproceedings{han2023systematic,
  author    = {Han, Yuyang and Ji, Xu and Wang, Zhiqiang and Zhang, Jianyi},
  year      = {2023},
  title     = {Systematic analysis of security and vulnerabilities in miniapps},
  booktitle = {{Proceedings of the 2023 {ACM} Workshop on Secure and Trustworthy Superapps}},
  pages     = {1--9}
}

@inproceedings{zhang2024minicat,
  author    = {Zhang, Zidong and Hou, Qinsheng and Ying, Lingyun and Diao, Wenrui and Gu, Yacong and Li, Rui and Guo, Shanqing and Duan, Haixin},
  year      = {2024},
  title     = {{MiniCAT}: Understanding and detecting cross-page request forgery vulnerabilities in mini-programs},
  booktitle = {{Proceedings of the 2024 {ACM} {SIGSAC} Conference on Computer and Communications Security}},
  pages     = {525--539}
}

@inproceedings{yang2022cross,
  author    = {Yang, Yuqing and Zhang, Yue and Lin, Zhiqiang},
  year      = {2022},
  title     = {Cross miniapp request forgery: Root causes, attacks, and vulnerability detection},
  booktitle = {{Proceedings of the 2022 {ACM} {SIGSAC} Conference on Computer and Communications Security}},
  pages     = {3079--3092}
}

@inproceedings{meng2023wemint,
  author    = {Meng, Shi and Wang, Liu and Wang, Shenao and Wang, Kailong and Xiao, Xusheng and Bai, Guangdong and Wang, Haoyu},
  year      = {2023},
  title     = {{WeMinT}: Tainting sensitive data leaks in {WeChat} mini-programs},
  booktitle = {{2023 38th {IEEE/ACM} International Conference on Automated Software Engineering ({ASE})}},
  pages     = {1403--1415}
}

@inproceedings{zhang2023don,
  author    = {Zhang, Yue and Yang, Yuqing and Lin, Zhiqiang},
  year      = {2023},
  title     = {Don't leak your keys: Understanding, measuring, and exploiting the {AppSecret} leaks in mini-programs},
  booktitle = {{Proceedings of the 2023 {ACM} {SIGSAC} Conference on Computer and Communications Security}},
  pages     = {2411--2425}
}

@inproceedings{wang2023taintmini,
  author    = {Wang, Chao and Ko, Ronny and Zhang, Yue and Yang, Yuqing and Lin, Zhiqiang},
  year      = {2023},
  title     = {{TaintMini}: Detecting flow of sensitive data in mini-programs with static taint analysis},
  booktitle = {{2023 {IEEE/ACM} 45th International Conference on Software Engineering ({ICSE})}},
  pages     = {932--944}
}

@inproceedings{wang2023one,
  author    = {Wang, Chao and Zhang, Yue and Lin, Zhiqiang},
  year      = {2023},
  title     = {One size does not fit all: Uncovering and exploiting cross platform discrepant {APIs} in {WeChat}},
  booktitle = {{32nd {USENIX} Security Symposium ({USENIX} Security 23)}},
  pages     = {6629--6646}
}

@inproceedings{Mohammadi,
  author    = {Mohammadi, Keyhan and Moradi, Arman and Atani, Reza Ebrahimi},
  year      = {2025},
  title     = {Security analysis of {MiniApps}: Vulnerabilities, exploits, and a tailored mitigation framework},
  booktitle = {{2025 15th International Conference on Computer and Knowledge Engineering ({ICCKE})}},
  pages     = {1--8},
  doi       = {10.1109/ICCKE68588.2025.11273836}
}

@misc{ferrari2026telegapper,
  author  = {Ferrari, Luca and Ceccato, Mariano and Verderame, Luca},
  year    = {2026},
  title   = {{TeleGapper}: On the {(Un)reliability} of privacy policies in {Telegram Mini Apps}},
  howpublished = {arXiv preprint arXiv:2608.13390}
}
